%% file: main.tex
\documentclass[aps,prl,twocolumn,amsmath,amssymb]{revtex4-2}
\usepackage{hyperref}
\usepackage{graphicx}
\usepackage{float}
\usepackage{color}
\usepackage{mathrsfs}
\usepackage{amsmath}

\newcommand{\jpsinum}{(10087\pm44)\times10^6}
\newcommand{\ee}{e^+e^-}
\newcommand{\jpsi}{J/\psi}
\newcommand{\pip}{\pi^+}
\newcommand{\pim}{\pi^-}
\newcommand{\piz}{\pi^0}
\newcommand{\Xim}{\Xi^-}
\newcommand{\Xip}{\bar{\Xi}^+}
\newcommand{\Lambdab}{\bar\Lambda}
\newcommand{\jpsixx}{\jpsi\to\Xi^{-}\bar{\Xi}^{+}}
\newcommand{\XimSemiDec}{{\Xi^{-}\rightarrow \Lambda e^- \bar\nu_{e}}}
\newcommand{\XipSemiDec}{{\bar{\Xi}^{+}\rightarrow \bar{\Lambda} e^+ \nu_{e}}}
\newcommand{\XimHadDec}{{\Xi^{-}\rightarrow \Lambda\pi^-}}
\newcommand{\XipHadDec}{{\bar{\Xi}^{+}\rightarrow \bar{\Lambda}\pi^+}}
\newcommand{\LambpChgDec}{\Lambda \rightarrow p \pi^-}
\newcommand{\LambmChgDec}{\bar{\Lambda} \rightarrow \bar{p} \pi^+}
\newcommand{\ppipi}{p\pi^-\pi^-}
\newcommand{\pbarpipi}{\bar{p}\pi^+\pi^+}
\newcommand{\ppi}{p\pi^-}
\newcommand{\pbarpi}{\bar{p}\pi^+}
\newcommand{\env}{e^-\bar{\nu}_e}
\newcommand{\MeVcc}{\mathrm{MeV/}c^2}
\newcommand{\GeVcc}{\mathrm{GeV/}c^2}

\begin{document}

\title{Measurements of the Absolute Branching Fraction of the Semileptonic Decay $\mathbf{\Xi^{-}\rightarrow \Lambda e^- \bar\nu_{e}}$ and the Axial Charge of the $\mathbf\Xi^{-}$ }

\author{\input{authorlist_2025-07-21}}

\date{\today}

\begin{abstract}
Using $\jpsinum$ $\jpsi$ events collected with the BESIII detector, we study the semileptonic decay $\XimSemiDec$ for the first time at an electron-positron collider. 
The absolute branching fraction is determined for the first time 
to be $(3.60\pm0.40_{\mathrm{stat}}\pm0.10_{\mathrm{syst}})\times10^{-4}$, 
which is 3.9 standard deviations below the world average. 
In addition, using an 11-dimensional angular analysis, the axial-vector to vector coupling $g_{av}$ is determined to be $0.18\pm0.07_{\mathrm{stat}}\pm0.01_{\mathrm{syst}}$. These results are used to test various SU(3)-flavour effective models. Under the SU(3)-flavour symmetry limit, the axial charge is found to be $g_A^H = 0.22\pm0.08_{\mathrm{stat}}\pm0.02_{\mathrm{syst}}$. 
Despite using only 5\% of the statistics of previous experiments, this analysis achieves a comparable precision for the axial charge.

\end{abstract}

\newcommand{\BESIIIorcid}[1]{\href{https://orcid.org/#1}{\hspace*{0.1em}\raisebox{-0.45ex}{\includegraphics[width=1em]{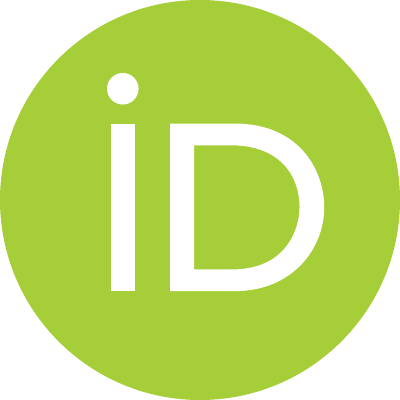}}}} 

\maketitle

The axial charge of a hyperon, $g_A^H$, is a fundamental observable in hadronic physics. Defined as the axial form factor at zero momentum transfer in hyperon semileptonic decay (HSD), it serves as an excellent probe for quantifying SU(3)-flavor symmetry breaking. 
It is also critical for determining the quark spin content of the proton, providing an important supplement to polarized deep inelastic scattering experiments~\cite{revmodphys.85.655}. Over six decades ago, Cabibbo introduced an SU(3)-flavor symmetry model for weak hadronic currents~\cite{physrevlett.10.531}, which universally parameterizes baryon axial charges via two coupling constants~\cite{annurev:/content/journals/10.1146/annurev.nucl.53.013103.155258}, $F$ and $D$. Remarkably, this framework has remained in good agreement with experimental data for decades, implying that SU(3) symmetry-breaking effects in HSD are intriguingly small.
This unusually small breaking effect has fueled extensive theoretical efforts. Predictions of axial charges have used the quark model~\cite{physrevd.82.014007,physrevd.51.2262,physrevc.97.055206,physrevc.105.065204,physrevd.35.934}, the large $N_c$ expansion \cite{physrevd.58.094028,physrevd.70.114036}, heavy baryon
chiral perturbation theory~\cite{Zhu:2002tn}, chiral soliton models~\cite{physrevc.92.035206}, and baryon chiral perturbation theory~\cite{physrevd.90.054502}, alongside lattice quantum chromodynamics computations~\cite{physrevd.108.034512}. 

Despite extensive theoretical studies of axial charges, experimental measurements remain scarce. Experimental measurements of $g_A^H$ rely on the axial-vector to vector coupling ratio in HSD, $g_A^H=g_{av}\cdot f_1$, where $f_1$ is the vector form factor at zero momentum transfer, which is protected from leading order SU(3)-breaking effects by the Ademollo-Gatto theorem~\cite{annurev:/content/journals/10.1146/annurev.nucl.53.013103.155258}. Precise measurements of $g_{av}$ are challenging because of the limited hyperon samples and the low branching fractions (BFs) of HSDs.

The BESIII experiment has recently accumulated a data sample of $\jpsinum$ $J/\psi$ events from $e^+e^-$ collisions~\cite{ablikim_2022}. The $J/\psi$ meson can decay into hyperon-antihyperon pairs with relatively large decay BFs of order $\sim10^{-3}$, offering a new platform to study the properties of hyperon decays~\cite{Li:2016tlt}.
Compared with traditional fixed-target experiments \cite{bristol-geneva-heidelberg-orsay-rutherford-strasbourg:1983jzt}, the hyperon pairs produced in the $J/\psi$ decays exhibit spin entanglement and polarization. Furthermore, benefiting from fully reconstructible decay chains and the joint angular distribution analysis method~\cite{physrevd.108.016011}, both the precision and sensitivity of such measurements are significantly improved.

In this Letter, we present a measurement of the decay $\XimSemiDec$, based on a sample of $\jpsinum$ $J/\psi$ events~\cite{ablikim_2022} collected by the BESIII detector. The corresponding absolute BF and the axial charge are reported.
Throughout this Letter, the $\Lambda$ hyperon is reconstructed via its charged decay mode $\LambpChgDec$, and charge conjugation is implied unless stated otherwise.

Details of the design and performance of the BESIII detector 
are provided in Ref.~\cite{ablikim2010345}. Simulated signal and background samples produced with a {\sc geant4}-based~\cite{agostinelli2003250} Monte Carlo (MC) software package  
are used to determine the detection efficiencies and estimate background contributions. The simulation models the beam energy spread and initial state radiation in the $e^+e^-$ annihilations with the generator {\sc kkmc}~\cite{jadach2000260}. 
The signal decay $\XimSemiDec$ and the dominant background decay $\XimHadDec$ are modeled according to Refs.~\cite{physrevd.108.016011, ablikim2022}, with decay parameters fixed to the latest measurements~\cite{ablikim2022,annurev:/content/journals/10.1146/annurev.nucl.53.013103.155258}. The inclusive MC sample includes both $J/\psi$ resonance production and continuum processes incorporated in {\sc
kkmc}~\cite{jadach2000260}. All particle decays are simulated using {\sc
evtgen}~\cite{evtgen} with branching fractions taken from the Particle Data Group (PDG)~\cite{physrevd.110.030001}
when available, or otherwise modeled with {\sc lundcharm}~\cite{Chen:2000+Yang:2014}. Final state radiation from charged final state particles is incorporated using the {\sc
photos} package~\cite{barberio:1990ms}.

To measure the BF of the decay $\Xi^-\to\Lambda e^{-}\bar{\nu}_{e}$, a double-tag (DT) technique is employed. In this method, $J/\psi\to\Xi^-\bar{\Xi}^+$ events are first selected by fully reconstructing a $\bar{\Xi}^+$ via its dominant decay mode $\bar{\Xi}^+\to\bar{\Lambda}\pi^+$. Such events are referred to as single-tag (ST) candidates. The signal decay $\Xi^-\to\Lambda e^{-}\bar{\nu}_{e}$ is then searched for in the system recoiling against the tagged $\bar{\Xi}^+$. Events in which both sides are successfully reconstructed are classified as DT candidates.
The BF of $\XimSemiDec$ decay is obtained using
\begin{equation}
\mathcal{B}(\XimSemiDec)=\frac{N_{\mathrm{DT}}}{N_{\mathrm{ST}}}\cdot \frac{\epsilon_{\mathrm{ST}}}{\epsilon_{\mathrm{DT}}}\cdot \frac{1}{\mathcal{B}(\Lambda\rightarrow p \pi^-)},
\label{eq:branch}
\end{equation}
\noindent
where $N_{\mathrm{DT}}$ and $N_{\mathrm{ST}}$ are the DT and ST yields in data, respectively, $\epsilon_{\mathrm{DT}}$ and  $\epsilon_{\mathrm{ST}}$ are the corresponding detection efficiencies,  
and $\mathcal{B}(\Lambda\rightarrow p \pi^-)$ is the BF of $\Lambda\rightarrow p \pi^-$ quoted from PDG~\cite{physrevd.110.030001}.

For the cascade decay $\jpsixx \to \Lambda \env {\Lambdab} \pip $, with subsequent decays $\LambpChgDec$ and $\LambmChgDec$, the event is characterized by eleven helicity angles  $\xi=(\theta_{\Xim},\theta_{\Lambda},\phi_{\Lambda},\theta_{e^-},\phi_{e^-},\theta_{p},\phi_{p},\theta_{\Lambdab},\phi_{\Lambdab},\theta_{\bar{p}},\phi_{\bar{p}})$, defined in the reference frame described in Ref.~\cite{physrevd.108.016011}, along with the squared momentum transfer $q^2$ of the intermediate $W$ boson. The joint angular amplitude of the cascade decay chain is given by
\begin{equation}
\mathcal{W}(\xi;\omega) =\sum_{\mu=0}^{3}\sum_{\bar\nu=0}^{3}\sum_{\mu'=0}^{3}\sum_{\bar\nu'=0}^{3}C_{\mu\bar\nu}B^{\Xim}_{\mu\mu'}A^{\Lambda}_{\mu'0}A^{\Xip}_{\bar\nu\bar\nu'}A^{\Lambdab}_{\bar\nu'0},
\label{eq3}
\end{equation}
where $\omega$ denotes a vector of nine global parameters $\omega=(\alpha_{\Psi}, \Delta\Phi, \alpha_{\bar\Xi}, \phi_{\bar\Xi}, \alpha_{\bar\Lambda}, g_{av}, g_{w}, g_{av2}, \alpha_{\Lambda})$.
Here, $\alpha_{\Psi}$ and $\Delta\Phi$ are production parameters of $\jpsi\to \Xi^-\bar{\Xi}^+$, while $\alpha_{\bar\Xi}$, $\alpha_{\bar\Lambda}$ and $\alpha_{\Lambda}$ are the decay asymmetry parameters of the corresponding hyperon nonleptonic decay.
The parameters $g_{av}$, $g_{w}$ and $g_{av2}$ are the axial-vector to vector coupling, weak-magnetic coupling and weak-electric coupling of HSD, respectively. Similar to $g_{av}$, $g_{w}$ is defined as $f_{2}/f_{1}$, and $g_{av2}$ is defined as $g_{2}/f_{1}$, where $f_{2}$ and $g_{2}$ are the weak-magnetism form factor and the weak-electricity form factor at zero momentum transfer, respectively. The formalism of the joint angular distribution, along with the definition of symbols in Eq.~\ref{eq3}  ($C_{\mu\nu}$, $A^{Y}_{\mu\nu}$ and $B^{Y}_{\mu\nu}$), is detailed in Ref.~\cite{physrevd.108.016011}.

To select events, good charged tracks detected in the main drift chamber (MDC) must satisfy $|\mathrm{{cos\theta}}|<$0.93, where $\theta$ is the polar angle with respect to the $e^{+}$ beam direction. Because of the kinematic separation between the protons and pions (electrons), charged tracks with momentum greater than 0.32 $\mathrm{GeV/}c$ are identified as (anti-)protons, while those with momentum less than 0.30 $\mathrm{GeV/}c$ are assigned as pions for ST, or as pions or electrons for DT. 
In the ST selection, to improve the purity, an additional particle identification (PID) is applied for anti-proton candidates.
The PID likelihoods, $\mathcal{L}_{p}$, $\mathcal{L}_{\pi}$ and $\mathcal{L}_{K}$, are calculated by combining the energy loss ($dE/dx$) information from the MDC and the time-of-flight measurements. The anti-proton candidates are required to satisfy $\mathcal{L}_{p}>\mathcal{L}_{\pi}$ and $\mathcal{L}_{p}>\mathcal{L}_{K}$. 
Events with at least an anti-proton and two positively charged pions are selected. The sequential decay $\XipHadDec\rightarrow \pbarpipi$ is reconstructed by a vertex fit \cite{ablikim2022,xu_min_2010}, which takes into account the flight paths of the hyperons. 
If there are multiple combinations, the one with the smallest $(M_{\pbarpi}-m_{\Lambdab})^{2}/\sigma_{\Lambdab}^2 + (M_{\pbarpipi}-m_{\Xip})^{2}/\sigma_{\Xip}^2$ is retained, where $M_{\pbarpipi (\pbarpi)}$ denotes the invariant mass of $\pbarpipi(\pbarpi)$ obtained from the vertex fit, $m_{\Xip(\Lambdab)}$ refers to the nominal mass of $\Xip(\Lambdab)$ quoted from PDG~\cite{physrevd.110.030001}, and $\sigma_{\Xip(\Lambdab)}$ represents the corresponding mass resolution.
The candidate events are further required to satisfy $|M_{\pbarpi}-m_{\Lambdab}|<6~\MeVcc$ and $|M_{\pbarpipi}-m_{\Xip}|<8~\MeVcc$. 
The decay lengths of the $\Xip$ and $\Lambdab$, obtained from the vertex fit, are required to be positive. Detailed MC studies indicate that the probability for a $\pip$ to be assigned to the wrong $\Lambdab$ or $\Xip$ decay is less than $0.1\%$, and thus  is negligible.

\begin{figure}[htbp]
\vspace{-0.1cm}
	\centering
	{
		\begin{minipage}{0.505\linewidth}
		\includegraphics[scale=0.22]{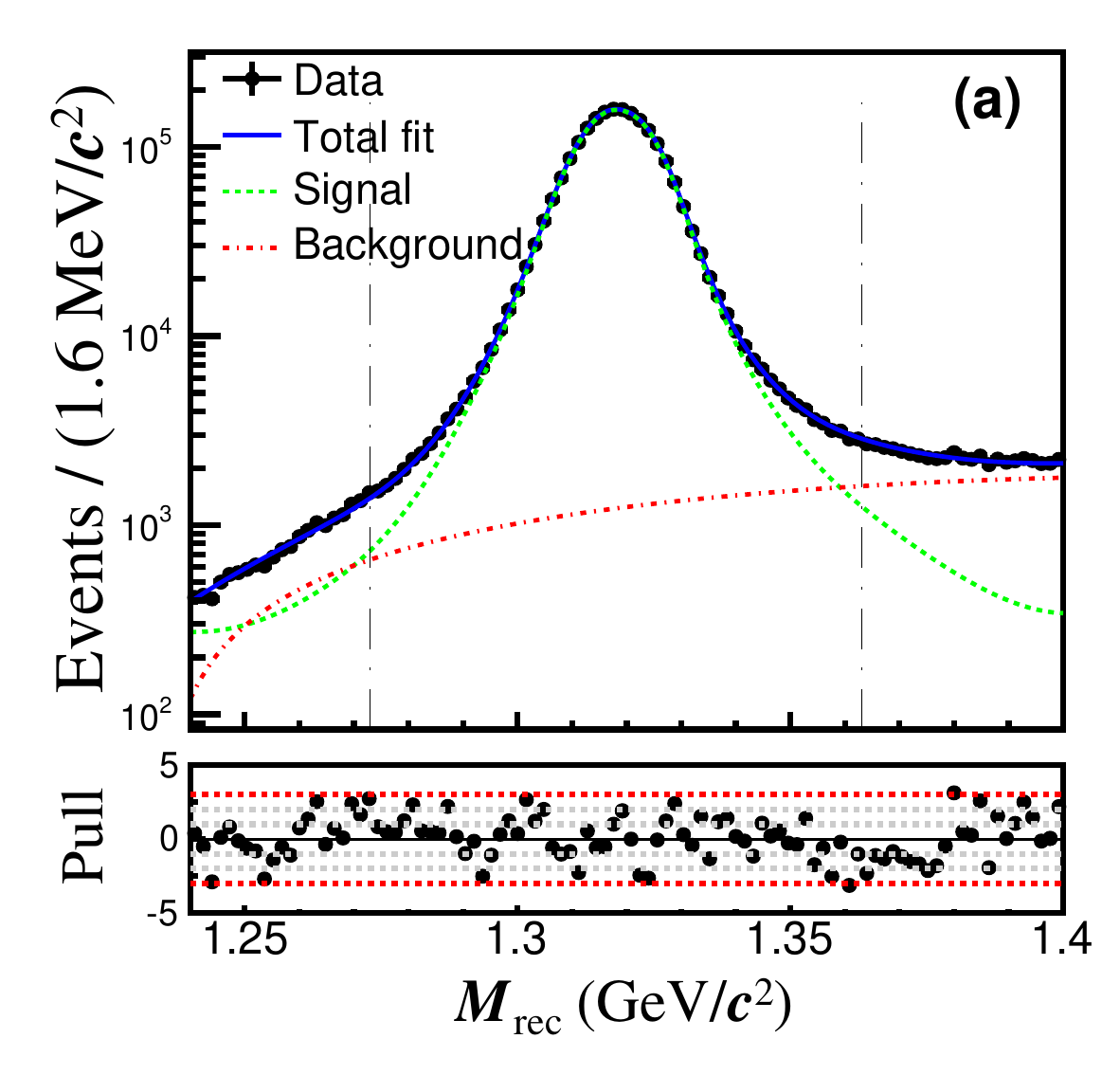}
		\end{minipage}
	}
 \hfill
 \hspace{-15pt}
	{
		\begin{minipage}{0.505\linewidth}
		\includegraphics[scale=0.22]{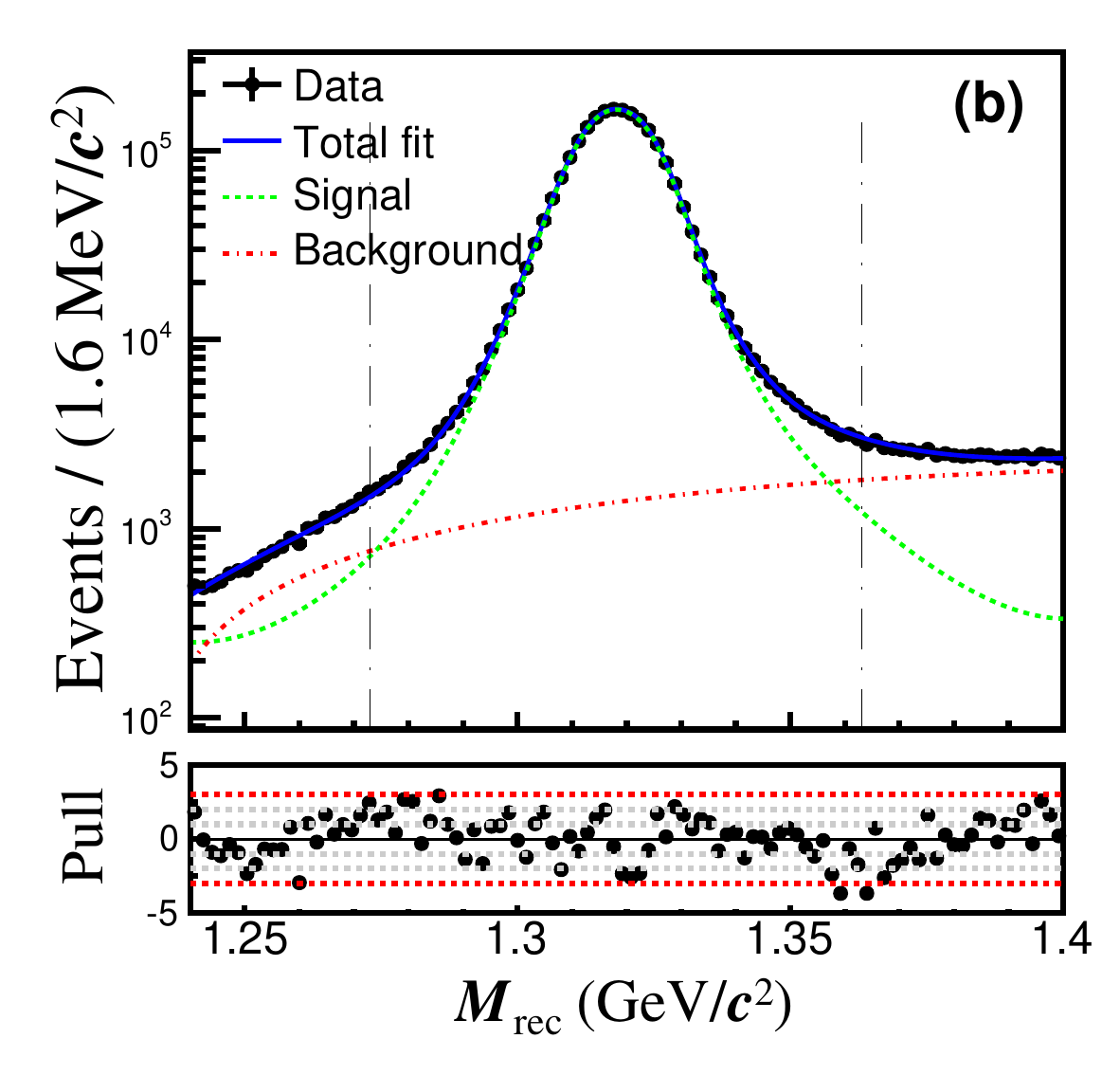}
		\end{minipage}
	}
  \setlength{\abovecaptionskip}{-3pt}
	\caption{Fits to the $M_{\mathrm{rec}}$ distributions of (a) ST $\Xip$ and (b) ST $\Xim$ candidate events. The points with error bars are data. The blue solid curves are the total fit results. The green dashed and red dot-dashed curves are the signal and background, respectively. The area between the two dashed lines represents the interval for extracting the ST signal. The red dashed line in the pull distribution marks the $3\sigma$ (3 standard deviation) deviation threshold. The two grey dashed lines in the pull distribution mark the $1\sigma$ and  $2\sigma$ deviation threshold.}
	\label{recfit}
\end{figure}

Finally, the ST yield is obtained from the mass distribution
 recoiling against the reconstructed $\bar{\Xi}^{+}$, defined as
\begin{equation}
M_{\mathrm{rec}}\equiv\sqrt{(E_{\mathrm{CM}}-E_{\bar\Lambda\pip})^2-|\vec{p}_{\bar\Lambda\pip}|^2},
\end{equation} where $E_{\mathrm{CM}}$ is the center-of-mass energy of the $e^+e^-$ system, and $E_{\bar\Lambda\pip}$ and $\vec{p}_{\bar\Lambda\pip}$ are the energy and momentum of the selected $\Xip$ candidate in the center-of-mass system, respectively. 
The $M_{\mathrm{rec}}$ distributions of the remaining events are shown in  Fig.~\ref{recfit} for the two charged conjugated modes individually, where signals are observed above a very low background.
Maximum likelihood fits to the $M_{\mathrm{rec}}$ distributions are performed to extract the ST yields. Using kernel density estimation (KDE), we obtain the signal MC shape, which is then convolved with a double-Gaussian function to account for discrepancies between data and MC, and used as the signal template. While the combinational background is described with a second-order Chebychev polynomial. No peaking background is found by studying the inclusive MC sample. To ensure the quality of ST reconstruction, the ST signal yields are 
obtained within the range $1.273 < M_{\mathrm{rec}} < 1.363~\GeVcc$, corresponding to three standard deviations.
The yields and the detection efficiencies determined from MC simulation are summarized in Table~\ref{table:thesignalyield}. It is worth noting that, as antiprotons are more prone to interact with the detector material compared with protons, the reconstruction efficiency of $\bar\Xi^+$ is lower than that of $\Xi^-$.

\begin{table}[htbp]
	\centering

	\caption{The values of $N_{\mathrm{ST}}$, $\epsilon_{\mathrm{ST}}(\%)$, $N_{\mathrm{DT}}$ and $\epsilon_{\mathrm{DT}}(\%)$, as well as results for the BF and $g_{av}$, obtained in the individual and simultaneous fits between the two charged conjugate modes. The first uncertainties are statistical and the second systematic, if present.}
	\label{table:thesignalyield}

   {
	\begin{tabular}{ c c c } \hline	\hline
	 Mode & ~~~~~$\XimSemiDec$ ~~~~~ & ~~~~~$\XipSemiDec$~~~~ \\ 	\hline

        $N_{\mathrm{ST}}$($10^{3}$) &2069.5 $\pm$ 1.8   & 2157.6 $\pm$ 1.9 \\ 

        $\epsilon_{\mathrm{ST}}(\%)$  & $28.71\pm0.02$ & $30.53\pm0.02$ \\
        $N_{\mathrm{DT}}$ & $49.4\pm8.5$ & $61.5\pm9.4$\\
        $\epsilon_{\mathrm{DT}}(\%)$  & $3.42\pm0.02$ & $3.37\pm0.02$ \\
        Indiv. BF$(10^{-4})$& $3.13\pm0.54$ & $4.04\pm0.62$ \\
        Simul. BF$(10^{-4})$ & \multicolumn{2}{c}{$3.60\pm0.40\pm0.10$}\\
        Indiv. $g_{av}$& $0.14\pm0.11$ & $-0.21\pm0.09$\\
        Simul. $|g_{av}|$ & \multicolumn{2}{c}{$0.18\pm0.07\pm0.01$}\\
        $g^H_{A}$ $(f_1=\sqrt{3/2})$ & \multicolumn{2}{c}{$0.22\pm0.08\pm0.02$}\\
    \hline
	\hline

	\end{tabular}
}
\end{table}

Starting with the sample of ST candidates, the signal decays $\XimSemiDec$ are reconstructed using the remaining tracks recoiling against the ST $\Xip$ candidates. 
At least one proton and two tracks with negative charge are required.
Similar to the ST reconstruction, the sequential decay $\XimSemiDec$ is reconstructed by a vertex fit. 
The proton and pion candidates are selected by requiring the minimum $|M_{\ppi}-m_{\Lambda}|$. The track with the minimum $\chi^{2}$ obtained from the primary vertex fit with the $\Lambda$ is assigned as the electron. Furthermore, the vertex fit is required to satisfy $\chi^{2}<20$ to ensure a good fit quality. The same requirements on $M_{\ppi}$ and the decay lengths of the $\Lambda$ and $\Xim$ are applied as in the ST selection.
Studies based on MC samples indicate that the dominant background arises from the nonleptonic decay $\XimHadDec$, due to its similar decay topology and much larger BF.
To suppress this background, a vertex fit identical to that used in ST $\Xip$ reconstruction is performed under the $\XimHadDec$ hypothesis, and candidates satisfying $|M_{\ppipi}-m_{\Xim}|<5.4~\MeVcc$, corresponding to three times the $\Xim$ mass resolution, are vetoed. 
This veto reduces the $\XimHadDec$ background by $96\%$, while retaining $88\%$ of the signal.
After this suppression, the background $\XimHadDec$ remains more than 150 times
larger than the expected signal, primarily due to misreconstructed pions from $\Xim$ decays caused by hard scattering in the MDC or secondary weak decays into muons.
To further suppress this background, a boosted decision tree (BDT)~\cite{therhaag:2011jh} classifier is adopted based on the tracking information to discriminate signal electrons from background pions. The BDT classifier is trained using the signal MC and dominant background MC samples. 
Input variables include the measured energy loss, the Z-scores of the measured energy loss relative to the expected values under different particle hypotheses, the track momentum, and the number of hits in the MDC.
If the track can reach the time-of-flight detector, the differences between the measured flight time and its expected value under different particle hypotheses are also incorporated. 
After applying the BDT selection, the signal efficiency is $66.3\%$, while $99.8\%$ of the dominant background is rejected.

After all the above requirements are imposed, background from the processes $\jpsi\to n\gamma\Lambda\Lambdab$(with $n\geq1$), including all $\Lambda\bar\Lambda$ intermediate states, becomes sizable.
The background arises when a photon converts into an electron-positron pair in the detector material, and the resulting positron is assigned to be a pion from the ST $\Xip$ decay.
To eliminate this background, the opening angle between the pion from the ST $\Xip$ and the electron from the signal side is required to be larger than $40^\circ$ ($\theta_{e^- \pi^+}>40^\circ$).

For the signal $\XimSemiDec$, the neutrino is not detected, and is characterized by the kinematic variable
\begin{equation}
U_{\mathrm{miss}}=E_{\mathrm{miss}}-c|\vec{p}_{\mathrm{miss}}|,
\end{equation}
where $E_{\mathrm{miss}}$ and $\vec{p}_{\mathrm{miss}}$ are the missing energy and momentum carried by the neutrino, respectively. These are calculated using
\begin{equation}
\begin{aligned}
E_{\mathrm{miss}} &= E_{\mathrm{beam}}-E_{\Lambda}-E_{e^-}, \\
\vec{p}_{\mathrm{miss}} &= \vec{p}_{\Xim}-\vec{p}_{\Lambda}-\vec{p}_{e^-},
\label{eq:pmiss}
\end{aligned}
\end{equation}
\noindent
where $E_{\mathrm{beam}}$ is the beam energy, $E_{\Lambda(e^-)}$ is the measured energy of $\Lambda(e^-)$, and $\vec{p}_{\Lambda(e^-)}$ is the corresponding momentum. To obtain better resolution, we use the constrained $\Xi^-$ momentum ($\vec{p}_{\Xim}$) given by 
 
\begin{equation}
\vec{p}_{\Xim} = -\frac{{\vec{p}_{\Xip}}}{|\vec{p}_{\Xip}|}\sqrt{E^2_{\mathrm{beam}}-m^2_{\Xip}c^4}/c.
\label{eq:pxi}
\end{equation}
All the quantities are evaluated in the center-of-mass frame of the $J/\psi$. The effects arising from the deflection of $\Xi^-$ and $\bar{\Xi}^+$ in the magnetic field have been corrected for all three-momenta.

\begin{figure}[htbp]
\vspace{-0.1cm}
	\centering
	{
		\begin{minipage}{0.505\linewidth}
		\includegraphics[scale=0.22]{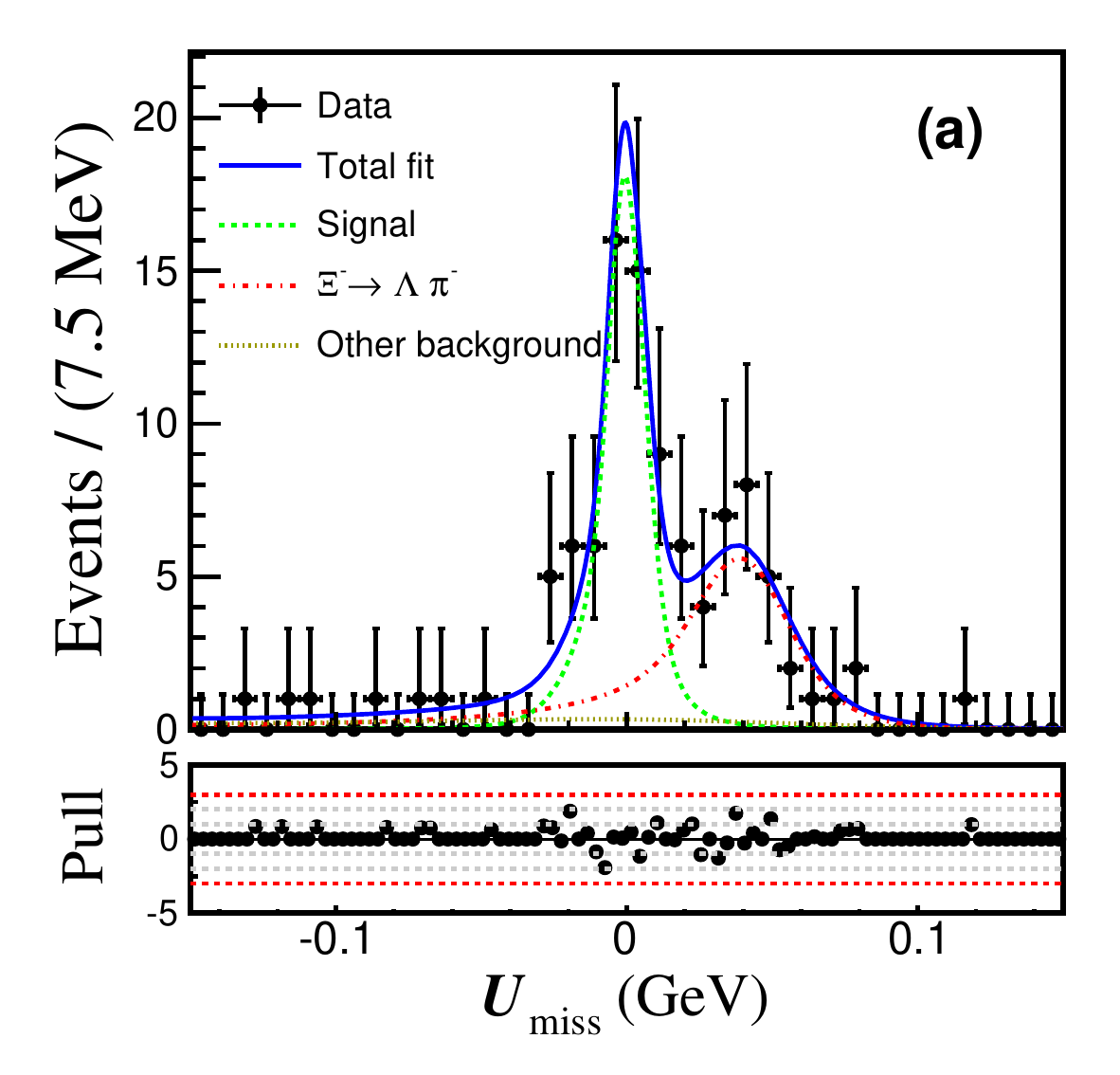}
		\end{minipage}
	}
  \hspace{-15pt}
	{
		\begin{minipage}{0.505\linewidth}
		\includegraphics[scale=0.22]{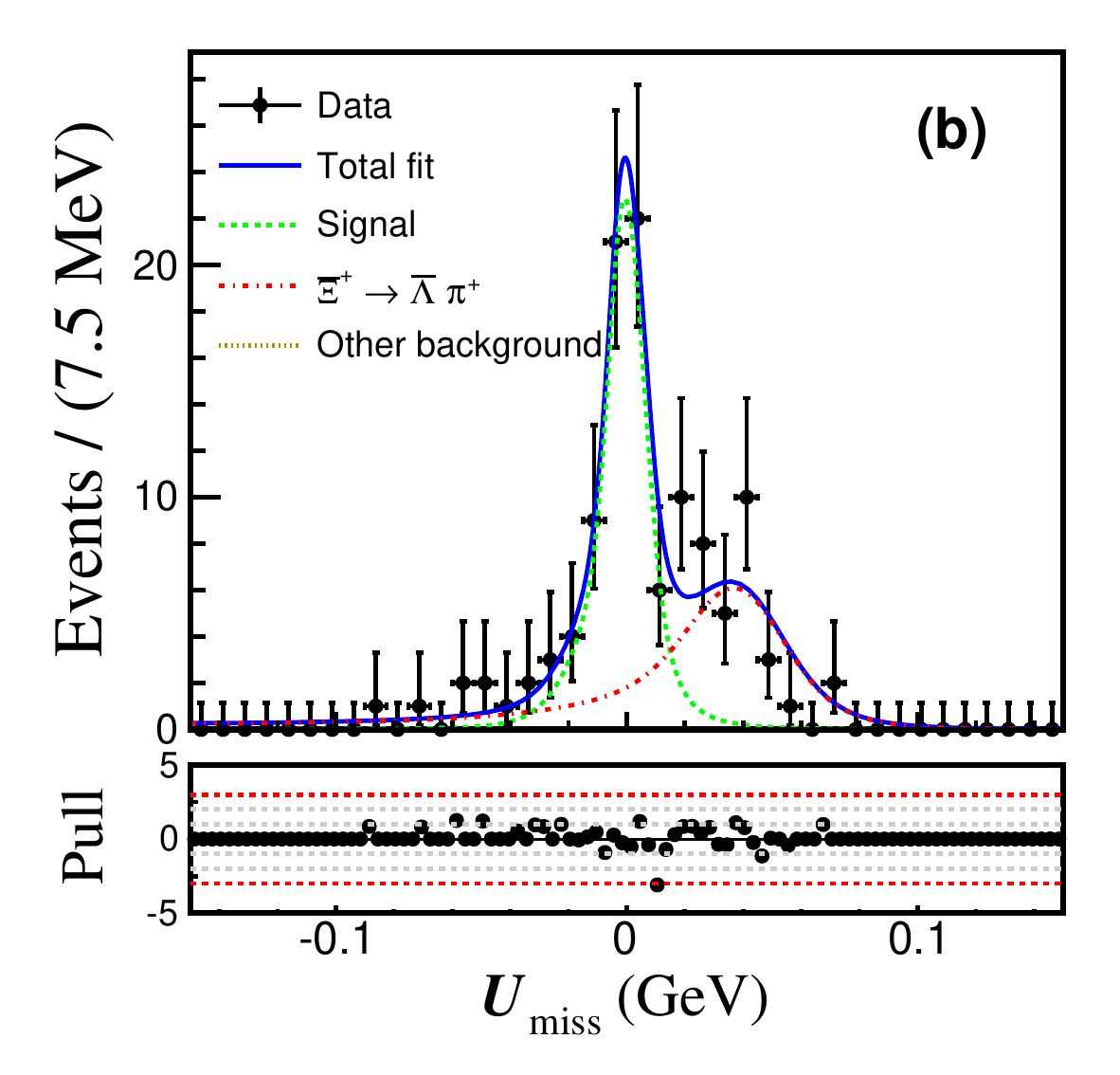}
		\end{minipage}
	}
   \setlength{\abovecaptionskip}{-3pt}
	\caption{Fits to the $U_{\mathrm{miss}}$ distributions for the (a) $\XimSemiDec$ and (b) $\XipSemiDec$ candidate events. The points with error bars are data. The blue solid curves are the total fit results. The green dashed curve is the signal.  The red dot-dashed and the grey dotted curves show the backgrounds from $\XimHadDec$($\XipHadDec$) and other sources, respectively. The red dashed line in the pull distribution marks the $3\sigma$ deviation threshold. The two grey dashed lines in the pull distribution mark the $1\sigma$ and  $2\sigma$ deviation threshold.}
 	\label{Umissfit}
\end{figure}

The $U_{\mathrm{miss}}$ distribution of the remaining candidates in data is shown in Fig.~\ref{Umissfit}. The signal is expected to peak around zero, while the dominant background $\XimHadDec$ is expected to peak around 0.04~GeV. There are also small background channels remaining, such as $\jpsi\rightarrow \Lambda\Lambdab\pip\pim$, $\jpsi\rightarrow \Sigma^{*+}\bar\Sigma^{*-}$. To extract the signal yield, an unbinned extended maximum likelihood fit is performed to the $U_{\mathrm{miss}}$ distribution. The signal is modeled by the KDE MC shape convolved with a Gaussian function, while the background $\XimHadDec$ is modeled by the KDE MC shape obtained from the exclusive MC sample. Other backgrounds are described by the inclusive MC simulated shape. 
The parameters of the Gaussian function used for convolution and all yields are left free in the fit. 
The fit curves are shown in Fig.~\ref{Umissfit} and the extracted signal yield, together with the detection efficiency from MC simulation, are summarized in Table~\ref{table:thesignalyield}.

According to Eq.~\ref{eq:branch}, the BFs of $\XimSemiDec$ and $\XipSemiDec$ are calculated and summarized in Table~\ref{table:thesignalyield}. The results from the two charge-conjugate modes are consistent within statistical uncertainties.
Therefore, a simultaneous fit to both charged-conjugate modes is performed, and the resulting BF is also shown in  Table~\ref{table:thesignalyield}.  

According to Eq.~\ref{eq3}, the axial-vector to vector coupling $g_{av}$ can be obtained by performing an unbinned maximum likelihood fit to the joint angular distribution of the surviving events $(N)$.  The likelihood is
\begin{equation}
\mathcal{L}(\omega)=\prod_{i=1}^N \frac{\mathcal{W}\left(\xi_i, \omega\right) \varepsilon\left(\xi_i\right)}{N(\omega)},
\label{eq:likelihood}
\end{equation}
\noindent
where $\mathcal{W}\left(\xi_i, \omega\right)$ is the probability density function as described in Eq.~\ref{eq3}, 
$\varepsilon\left(\xi_i\right)$ is the corresponding detection efficiency,
and $N(\omega)=(1/M)\sum_{i=1}^{M}[\mathcal{W}\left(\xi_i, \omega\right)/\mathcal{W}\left(\xi_i, \omega_{\rm gen}\right)]$ is the normalization factor evaluated with a signal MC sample of $M$ surviving events generated with parameters $\omega_{\rm gen}$. To further improve the purity of the data sample, only events within the region $-0.03<U_{\mathrm{miss}}<0.02~\mathrm{GeV}$ are used. A total of 120 events, including $100\pm11$ signal events from the two charge-conjugate modes, are included in the fit. 
In Eq.~\ref{eq:likelihood}, the background component is not included in $\mathcal{W}\left(\xi_i, \omega\right)$. Therefore, the background contribution $\mathcal{L}_{\mathrm{bkg}}(\omega)$ is evaluated with a MC-simulated background sample, and the fit is performed by minimizing the log likelihood function $S=-(\ln\mathcal{L}(\omega)-\ln\mathcal{L}_{\mathrm{bkg}}(\omega))$.

\begin{figure}[htbp]
\vspace{-0.1cm}
	\centering
	{
		\begin{minipage}{0.505\linewidth}
		\includegraphics[scale=0.22]{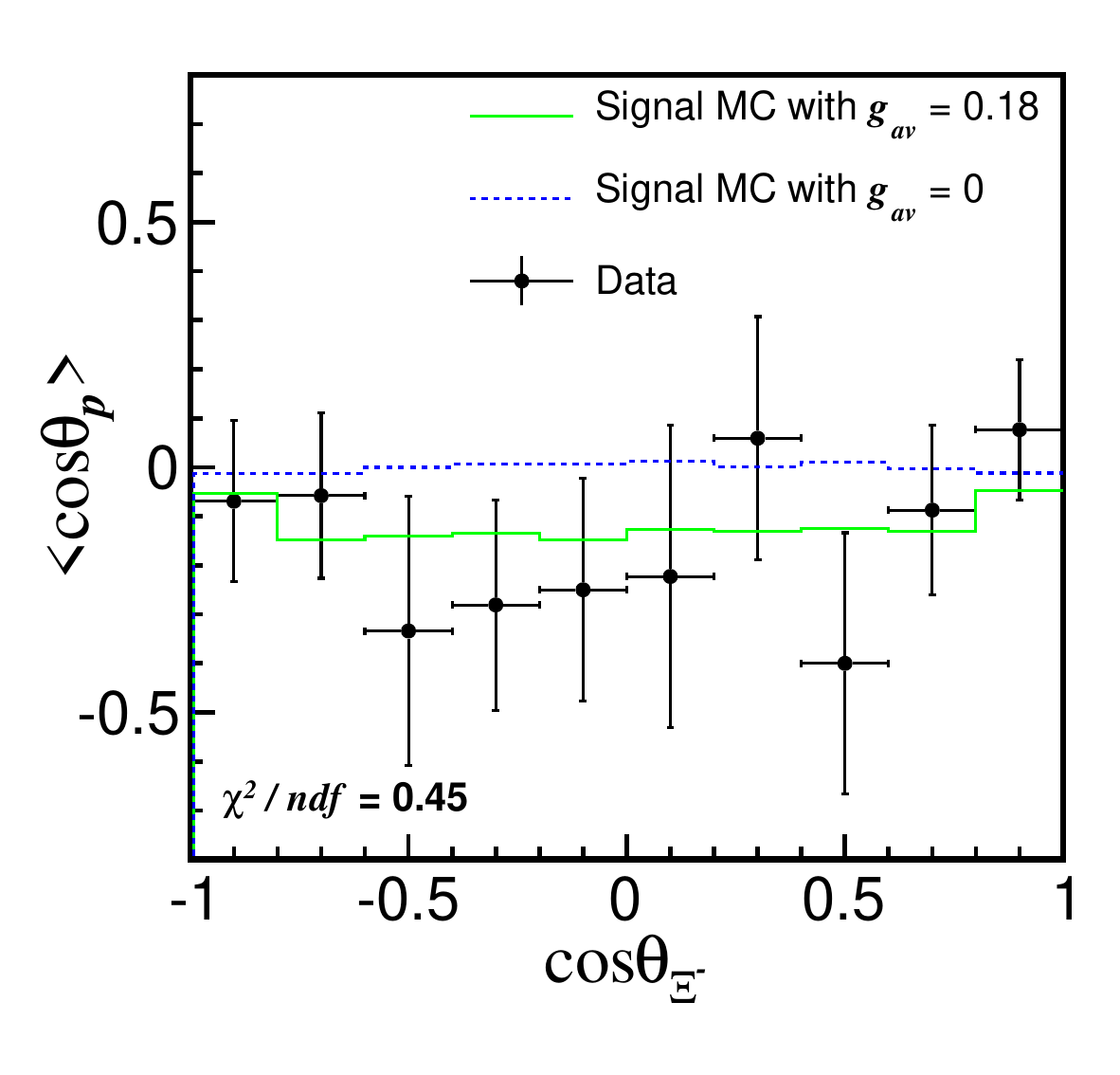}
		\end{minipage}
	}
  \hspace{-15pt}
	{
		\begin{minipage}{0.505\linewidth}
		\includegraphics[scale=0.22]{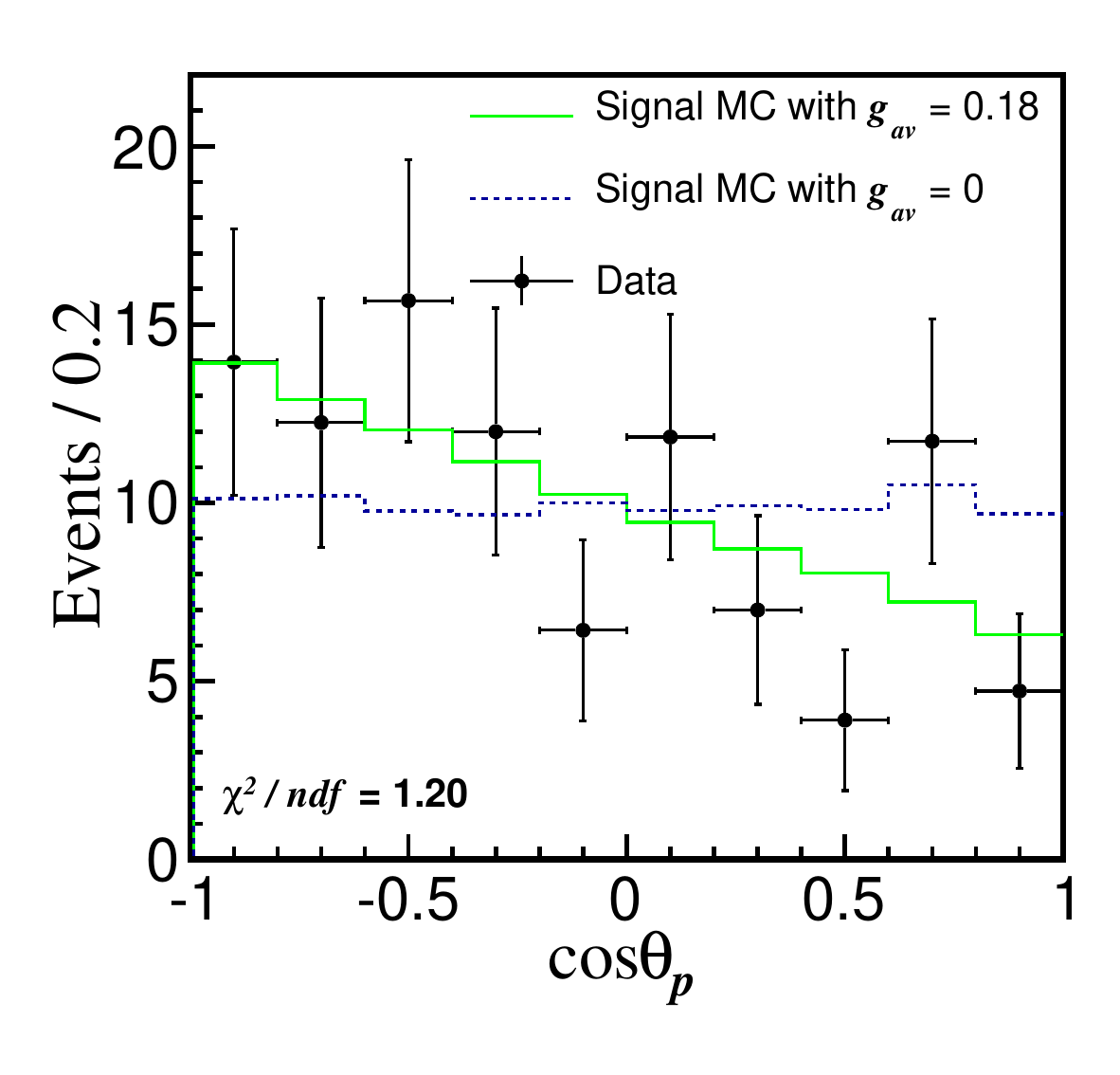}
		\end{minipage}
	}
    \setlength{\abovecaptionskip}{-3pt}
	\caption{The moment and $\mathrm{cos}\theta_{p}$ distributions. The black points with error bars are data. The green solid curves show the MC simulation based on the fit results.
    The blue dashed curves represent the MC simulation with $g_{av}=0$.}
 	\label{moments}
\end{figure}

The fits are performed for the two charge-conjugate modes and using a profile likelihood approach. The production parameters and decay asymmetry parameters are treated as float nuisance parameters, and multi-dimensional Gaussian contributions are added to the original log-likelihood function according to their central values and covariance matrix~\cite{ablikim2022}. The semileptonic decay form factors are fixed to $g_{w}=0.0846$ and $g_{av2}=0$ from Ref.~\cite{annurev:/content/journals/10.1146/annurev.nucl.53.013103.155258}, with only $g_{av}$ floated. The absolute values of $g_{av}$ obtained from the charge-conjugated modes are consistent with each other as summarized in Table~\ref{table:thesignalyield}.
A simultaneous fit assuming the same magnitude but opposite sign of $g_{av}$ for  the two charge-conjugate channels is also performed using the same approach, and the result is also shown in Table~\ref{table:thesignalyield}. 
To visualize the effect of $g_{av}$, a moment is calculated using $m=10$ intervals in the $\mathrm{cos}\theta_{\Xi^-}$ distribution as
\begin{align}
    \langle \mathrm{cos}\theta_{p}\rangle_k=\frac{m}{N}\sum_{i=1}^{N_k}\mathrm{cos}\theta^{i}_{p},
\end{align}
where $N_{k}$ is the number of events in the $k$-th $\mathrm{cos}\theta_{\Xi^-}$ interval. $\theta_{\Xi^-}$ and $\theta_{p}$ are the helicity angles of $\Xi^-$ and $p$, respectively, as defined in Ref.~\cite{physrevd.108.016011}. The $\mathrm{cos}\theta_{p}$ can also clearly demonstrate the effect of a non-zero $g_{av}$. Figure~\ref{moments} shows a comparison of moments and $\mathrm{cos}\theta_{p}$ between data and the MC projection based on the fit, where the background contributions in data have been subtracted. Good agreement between data and MC is observed.

The systematic uncertainties in the BF measurements are categorized into those from detection efficiency and those from the fit procedure. Since the DT method is applied, only the effects on the signal side are considered. All percentage values quoted below are relative uncertainties. In this analysis, the process $\jpsixx\to\Lambda\pim\Lambdab\pip$ is the dominant background source, while also serving as a high-purity control sample for the evaluation of core systematic uncertainties in the measurement. Uncertainties associated with $\pi^+(\pi^-)$ tracking efficiency (0.1\%), $p(\bar{p})$ tracking efficiency (0.1\%), and vertex fitting selection requirements (0.7\%) — including the $\Lambda(\bar{\Lambda})$ invariant mass window and decay length criteria for $\Lambda(\bar{\Lambda})$ and $\Xi^-(\bar{\Xi}^+)$ candidates  — are quantified using this control sample. The uncertainties of the tracking (0.2\%) and the BDT method (0.5\%) efficiencies for electrons are studied with the control samples $\jpsi\to\gamma \ee$ and $\jpsi\to\pip\pim\piz\to\pip\pim\gamma \ee$, respectively. It is worth noting that the control sample $\jpsi\to\pip\pim\piz\to\pip\pim\gamma \ee$ is divided into two subsets according to the availability of TOF information, and separate validation is performed for the BDT classifiers with and without TOF information inputs. The uncertainties due to $|M_{\ppipi}-m_{\Xim}|>5.4~\MeVcc$ (2.2\%) and $\theta_{e^- \pi^+}>40^\circ$ (0.4\%) are studied with the same control sample $\jpsixx\to\Lambda\pim\Lambdab\pip$. The uncertainties of the ST and DT yields are estimated to be $0.6\%$ and $2.5\%$, respectively, by changing the fit parameters and fit ranges. The uncertainties associated with the efficiency of the signal MC model (1.2\%) are estimated with the alternative MC samples generated by varying the model parameters according to central values and the uncertainties quoted in Ref.~\cite{ablikim2022}, Ref.~\cite{annurev:/content/journals/10.1146/annurev.nucl.53.013103.155258}, and the $g_{av}$ result of this work. Lastly, the uncertainty from the branching fraction of $\Lambda\to p \pi^-$ is 0.8\%. Further details on systematic uncertainties are provided in the Supplemental Material~\cite{Supp}. Assuming all the sources are uncorrelated, the total systematic uncertainty in the BF measurement is $3.8\%$, obtained by summing all contributions in quadrature.

The systematic uncertainties of the axial-vector to vector coupling $g_{av}$ include contributions from the fit process and event selection. All uncertainty values reported below are absolute uncertainties. The uncertainties related to the fit process are estimated with alternative fits, in which the number of background events is varied by $\pm1\sigma$. The MC-data discrepancies associated with event selections include those related to the tracking and PID of protons, the tracking of pions, the vertex fit reconstruction of $\Lambda$, as well as the tracking and BDT of electrons are studied with the same control samples as utilized in the  BF uncertainties. 
Alternative fits incorporating the above differences into the MC sample are performed, and the resulting variations of $g_{av}$ are taken as the corresponding uncertainties. The uncertainties of other selections are studied by varying the selection criteria around their nominal values and repeating the fit process. Further details are provided in the Supplemental Material \cite{Supp}.
Assuming all the sources are uncorrelated, the total absolute systematic uncertainty in $g_{av}$ is $0.01$, obtained by summing up all the above contributions in quadrature.

\begin{figure}[htbp]
\vspace{-0.1cm}
	\centering
	{
		\begin{minipage}{1\linewidth}
		\includegraphics[scale=0.4]{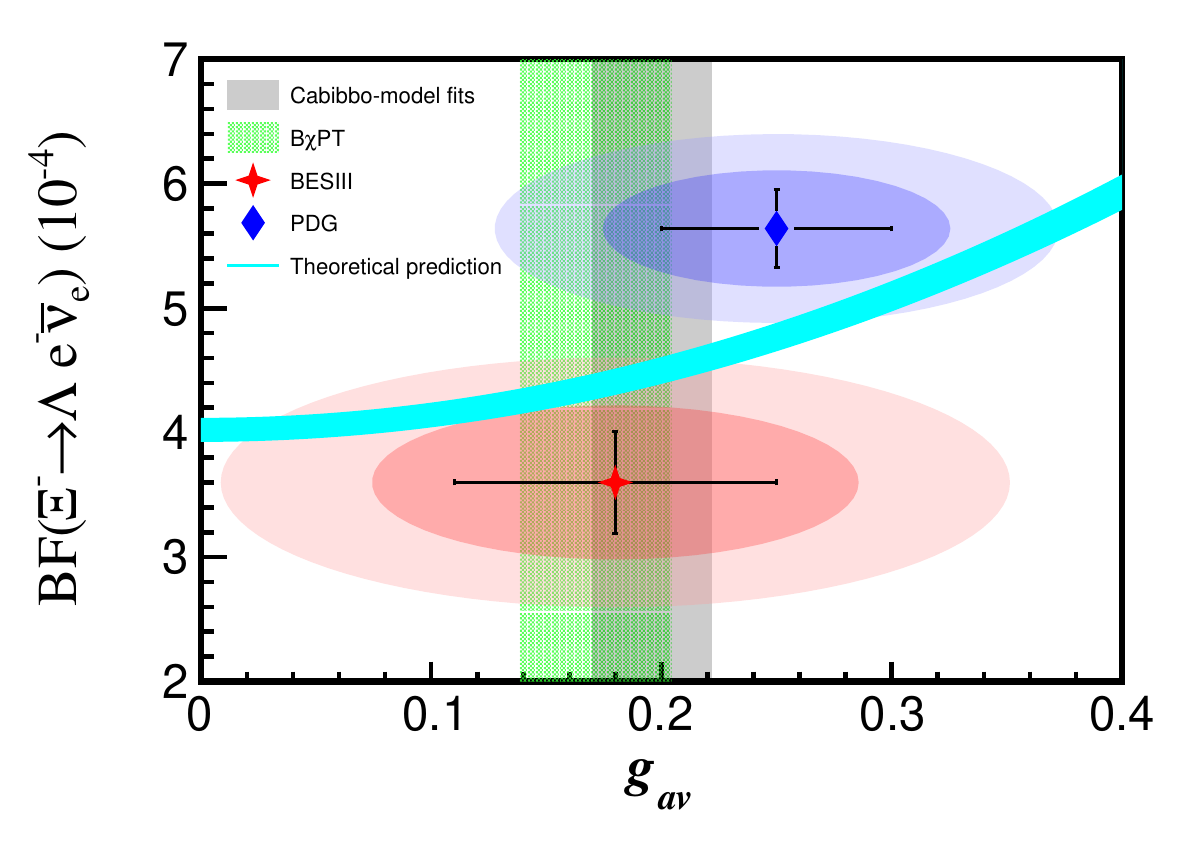}
		\end{minipage}
	}
     \setlength{\abovecaptionskip}{-3pt}
	\caption{Distribution of BF versus $g_{av}$ of the $\XimSemiDec$ decay. The red star denotes the results measured by this work and the red contours correspond to the $68\%/95\%$ confidence level of the results. The blue diamond represents the world average values~\cite{bristol-geneva-heidelberg-orsay-rutherford-strasbourg:1983jzt} of the BF and $g_{av}$, with the blue contours corresponding to the $68\%/95\%$ confidence level regions. The cyan line shows the predicted BF as a function of $g_{av}$, cited from Ref.~\cite{physrevd.108.016011}, where the input $|V_{us}|$ is set to the global fit value of the CKM matrix~\cite{physrevd.110.030001}. The grey box shows $g_{av}$ from the Cabibbo model global fit~\cite{annurev:/content/journals/10.1146/annurev.nucl.53.013103.155258}. The green box shows the $g_{av}$ prediction from baryon chiral perturbation theory (B$\chi$PT)~\cite{physrevd.90.054502}.}

 	\label{gav}
\end{figure}

In summary, using a data sample of $(10087\pm44)\times10^{6}$ $\jpsi$ events collected with the BESIII detector, the HSD $\XimSemiDec$ is studied at an electron-positron
 collider for the first time. The absolute BF is determined to be $(3.60\pm0.40_{\rm stat}\pm0.10_{\rm syst})\times10^{-4}$ for the first time, which is lower than the world average~\cite{bristol-geneva-heidelberg-orsay-rutherford-strasbourg:1983jzt} value by 3.9$\sigma$.
 The axial-vector to vector coupling $g_{av}$, is obtained as $0.18\pm0.07_{\rm stat}\pm0.01_{\rm syst}$, consistent with the world average value~\cite{bristol-geneva-heidelberg-orsay-rutherford-strasbourg:1983jzt} within uncertainty and in good agreement with prediction of chiral perturbation theory~\cite{physrevd.90.054502}, as illustrated in Fig.~\ref{gav}. Taking $f_1=\sqrt{3/2}$ within the SU(3)-flavor symmetry limit, the axial charge of the $\Xim$ is determined to be $g_A^H=0.22\pm0.08_{\rm stat}+0.02_{\rm syst}$. For the first time in over four decades, this measurement updates the results for the axial charge of the $\Xi^-$ HSD. These results provide critical inputs for studying SU(3)-flavor symmetry breaking effects in baryons and the quark spin content of the proton.
 Notably, a comparable precision on the axial-vector to vector coupling, $g_{av}$, is achieved using only one-twentieth of the data size employed in previous experiments~\cite{bristol-geneva-heidelberg-orsay-rutherford-strasbourg:1983jzt}. Such a remarkable improvement in measurement sensitivity arises from two key factors: the use of the advanced joint angular-distribution fit method~\cite{physrevd.108.016011}, which maximizes the exploitation of data compared to the original single-kinematic-variable fit approach~\cite{bristol-geneva-heidelberg-orsay-rutherford-strasbourg:1983jzt}, and the quantum-entangled production of polarized hyperon–antihyperon pairs at electron–positron colliders. This work is the first exploitation of polarization and quantum entanglement in cascade baryon semi-leptonic decays. This methodology is portable and can be readily extended to other decays at BESIII~\cite{besiiicollaboration2025determinationckmmatrixelement} and to future facilities, including PANDA~\cite{panda:2020zwv} and proposed Super Tau-Charm Factories~\cite{achasov:2023gey,barnyakov:2020vob}.
 
The BESIII Collaboration thanks the staff of BEPCII (https://cstr.cn/31109.02.BEPC) and the IHEP computing center and the supercomputing center of the University of Science and Technology of China (USTC) for their strong support. We thank Ruixiang Shi for his insightful theoretical discussions. This work is supported in part by National Key R\&D Program of China under Contracts Nos. 2023YFA1606000, 2023YFA1606704, 2023YFA1609400; National Natural Science Foundation of China (NSFC) under Contracts Nos. 11635010, 11935015, 11935016, 11935018, 12025502, 12035009, 12035013, 12061131003, 12105276, 12122509, 12192260, 12192261, 12192262, 12192263, 12192264, 12192265, 12221005, 12225509, 12235017, 12361141819; the Chinese Academy of Sciences (CAS) Large-Scale Scientific Facility Program; the Strategic Priority Research Program of Chinese Academy of Sciences under Contract No. XDA0480600; CAS under Contract No. YSBR-101; Joint Large-Scale Scientific Facility Funds of the NSFC and CAS under Contracts Nos. U2032111; 100 Talents Program of CAS; CAS Youth Team Program under Contract No. YSBR-101; The Institute of Nuclear and Particle Physics (INPAC) and Shanghai Key Laboratory for Particle Physics and Cosmology; ERC under Contract No. 758462; German Research Foundation DFG under Contract No. FOR5327; Istituto Nazionale di Fisica Nucleare, Italy; Knut and Alice Wallenberg Foundation under Contracts Nos. 2021.0174, 2021.0299; Ministry of Development of Turkey under Contract No. DPT2006K-120470; National Research Foundation of Korea under Contract No. NRF-2022R1A2C1092335; National Science and Technology fund of Mongolia; Polish National Science Centre under Contract No. 2024/53/B/ST2/00975; STFC (United Kingdom); Swedish Research Council under Contract No. 2019.04595; U. S. Department of Energy under Contract No. DE-FG02-05ER41374; Beijing Natural Science Foundation of China (BNSF) under Contract No. IS23014.

\bibliographystyle{apsrev4-2.bst}
\bibliography{Reference}

\end{document}

%% file: authorlist_2025-07-21.tex
M.~Ablikim$^{1}$\BESIIIorcid{0000-0002-3935-619X},
M.~N.~Achasov$^{4,c}$\BESIIIorcid{0000-0002-9400-8622},
P.~Adlarson$^{81}$\BESIIIorcid{0000-0001-6280-3851},
X.~C.~Ai$^{86}$\BESIIIorcid{0000-0003-3856-2415},
R.~Aliberti$^{38}$\BESIIIorcid{0000-0003-3500-4012},
A.~Amoroso$^{80A,80C}$\BESIIIorcid{0000-0002-3095-8610},
Q.~An$^{77,63,\dagger}$,
Y.~Bai$^{61}$\BESIIIorcid{0000-0001-6593-5665},
O.~Bakina$^{39}$\BESIIIorcid{0009-0005-0719-7461},
Y.~Ban$^{49,h}$\BESIIIorcid{0000-0002-1912-0374}
% ...
\\
\vspace{0.2cm}
(BESIII Collaboration)\\
\vspace{0.2cm} {\it
$^{1}$ Institute of High Energy Physics, Beijing 100049, People's Republic of China\\
% ...
$^{4}$ Budker Institute of Nuclear Physics SB RAS (BINP), Novosibirsk 630090, Russia\\
% ...
$^{38}$ Johannes Gutenberg University of Mainz, Johann-Joachim-Becher-Weg 45, D-55099 Mainz, Germany\\
$^{39}$ Joint Institute for Nuclear Research, 141980 Dubna, Moscow region, Russia\\
% ...
$^{49}$ Peking University, Beijing 100871, People's Republic of China\\
% ...
$^{61}$ Southeast University, Nanjing 211100, People's Republic of China\\
% ...
$^{63}$ State Key Laboratory of Particle Detection and Electronics, Beijing 100049, Hefei 230026, People's Republic of China\\
% ...
$^{77}$ University of Science and Technology of China, Hefei 230026, People's Republic of China\\
% ...
$^{80}$ University of Turin and INFN, (A)University of Turin, I-10125, Turin, Italy; (B)University of Eastern Piedmont, I-15121, Alessandria, Italy; (C)INFN, I-10125, Turin, Italy\\
$^{81}$ Uppsala University, Box 516, SE-75120 Uppsala, Sweden\\
% ...
$^{86}$ Zhengzhou University, Zhengzhou 450001, People's Republic of China\\
\vspace{0.2cm}
$^{\dagger}$ Deceased\\
% ...
$^{c}$ Also at the Novosibirsk State University, Novosibirsk, 630090, Russia\\
% ...
$^{h}$ Also at State Key Laboratory of Nuclear Physics and Technology, Peking University, Beijing 100871, People's Republic of China\\
% ...
}